\documentclass[prd,twocolumn,showpacs,preprintnumbers,superscriptaddress,floatfix,nofootinbib]{revtex4-1}

\usepackage{amsfonts}
\usepackage{multirow}
\usepackage{subfigure}

\usepackage{graphicx,,booktabs}
\usepackage{amssymb}
\usepackage{amsmath,txfonts,ulem}
\usepackage{graphicx}
\usepackage{dcolumn}
\usepackage{color}
\usepackage{bm}
\usepackage{ulem}
\usepackage[colorlinks,
citecolor=blue,
anchorcolor=green,
menucolor=orange,
linkcolor=blue,
filecolor=red,
runcolor=pink,
urlcolor=blue,
frenchlinks=red]{hyperref}

\allowdisplaybreaks[4]

\begin{document}

	\title{\boldmath   Gravitational form factors of the proton from near-threshold
		vector meson photoproduction }
	
	\author{Xiao-Yun Wang}
	\email{xywang@lut.edu.cn}
	\affiliation{Department of physics, Lanzhou University of Technology,
		Lanzhou 730050, China}
	\affiliation{Lanzhou Center for Theoretical Physics, Key Laboratory of Theoretical Physics of Gansu Province, and Key Laboratory of Quantum Theory and Applications of MoE, Lanzhou University, Lanzhou, Gansu 730000, China}
	
	\author{ Fancong Zeng}
	\email{ zengfc@ihep.ac.cn}
	\affiliation{ Institute of High Energy Physics, Chinese Academy of Sciences, Beijing 100049, China}
	\affiliation{ University of Chinese Academy of Sciences, Beijing 100049, China}
	
	\author{ Jiyuan Zhang}
	\email{ jyuanzhang@yeah.net}
	\affiliation{Department of physics, Lanzhou University of Technology,
		Lanzhou 730050, China}
	\date{\today}
	\begin{abstract} 
		We embark on a systematical analysis of the quark and gluon gravitational form factors (GFFs) of the proton, by connecting energy-momentum tensor (EMT) and the near-threshold vector meson photoproduction (NTVMP). 
		Concretely, 
		the quark contributions of GFFs are determined   by global fitting the cross section of the lightest vector meson $\rho^0$ photoproduction. 
		Combined with the gluon GFFs achieved from heavy quarkonium $J/\psi$ photoproduction data, the complete GFFs are obtained and compared with the  experimental results and Lattice QCD determinations. In addition, we use the Resonances Via Pad\'{e} (RVP) method based on the Schlessinger Point Method (SPM) to obtain a model-independent quark $D$-term distribution by direct analytical continuation of Deep Virtual Compton Scattering (DVCS) experimental data. If errors are considered, the results obtained by RVP are basically consistent with those obtained by NTVMP. Moreover, the comprehensive information on GFFs  helps us to uncover the mass distribution and mechanical properties inside the proton.
		This work is not only an important basis for delving the proton enigmatic properties, % unraveling the secrets of the proton  internal nature
		but also have significance theoretical guiding for future  JLab   and EICs experimental measurements.
		
	\end{abstract}
	
	\maketitle 
	\section{ Introduction} 
	The structure of hadrons, the bound states of strong interactions, is conveniently probed by exploring the other fundamental forces: electromagnetic, weak, and gravitational interactions. For the proton's overall properties, both electromagnetic and weak form factors can be well determined, while our understanding on gravitational form factors
	(GFFs) is inadequate.  As the gravity couples to matter via the energy-momentum tensor (EMT), the distributions of energy, angular momentum, and various mechanical properties can be obtained from the matrix elements of EMT \cite{Kobzarev:1962wt,Pagels:1966zza}. These properties are encoded in the GFFs.  Concretely, 
	the GFFs provide critical information about the fundamental properties: mass $M$, spin $J$, and the Druck term ($D$ term), and offer direct access to the proton's internal structure  \cite{Kharzeev:2021qkd,Polyakov:2018zvc,Polyakov:2002yz}.  %including its mass, spin and mechanical properties of the proton
	The sum GFFs, which are divided into quark and gluon contributions, are measurable quantities defined purely for the total system, providing information on the quantum origin of mass and the internal dynamics of the proton.
	
	The mass and pressure distributions in the proton are constructed from the $A$-term and $D$-term GFFs. However, besides the precise measurement of the proton mass and the quark $D$-term extracted from deeply virtual Compton scattering (DVCS) experiment \cite{Burkert:2018bqq}, the complete information on the proton GFFs is very meager. Recently, one study obtained the $D$-term under the assumptions of the quark and gluon $D$-term is the same, as there is no information on the gluon $D$-term from experiment \cite{Burkert:2018bqq}. On the other hand, the quark and gluon contributions to $D$-term have been researched using lattice quantum chromodynamics (LQCD) \cite{Shanahan:2018nnv,Shanahan:2018pib,Pefkou:2021fni,Hackett:2023rif}, enhancing our understanding of the proton GFFs. To date, the exact value of $D$-term,
	which varies with the squared momentum transfer $t$, remains a subject of ongoing research.
	
	To address this problem, the near-threshold vector meson photoproduction (NTVMP) process
	should be considered as a indispensable way. This is because in the indirect measurement of GFFs in exclusive processes, besides DVCS, $\gamma \gamma^*$ reaction, time-like Compton scattering and  double DVCS \cite{Burkert:2023wzr}, the  vector meson production experiment are regarded as an advantageous part and abundant experiment data have been collected   by comparison.  Actually, 
	the proton internal nature can be explored by using a photon as a probe in the $ \gamma p\to V p$ process, with the same vector quantum numbers as a photon and the vector meson which is $J^{PC}=1^{--}$,   such as the proton mass radius \cite{Wang,Kharzeev:2021qkd,Wang:2022vhr}, the trace anomaly contribution to the proton mass \cite{Wang:2019mza,Ji:2021pys,Ji:2021mtz,Wang:2022tzw}, as well as the quarkonium–proton scattering length \cite{Strakovsky:2014wja,Strakovsky:2020uqs,Strakovsky:2019bev,Strakovsky:2021vyk,Wang:2022xpw}, etc. Moreover, the near-threshold heavy quarkonium photoproduction offers a superior path to access the gluon GFFs \cite{Wang:2022ndz,Kharzeev:2021qkd,Guo:2021ibg,Hatta:2019lxo,Hatta:2018ina,Mamo:2019mka,Ji:2020bby,Sun:2021gmi,Duran:2022xag,Guo:2021ibg}. 
	One mainly reason is that the scalar gluon operator is dominant in the production amplitude of heavy quarkonium (such as $J/\psi$), and sensitive to the gluonic structure of the proton \cite{Kharzeev:2021qkd}. 
	Actually, the light vector meson (such as $\rho^0$) photoproduction
	mainly reflects the quark part of the GFFs \cite{Wang},
	since the exchange of a scalar quark-antiquark pair is not suppressed and far exceeds the contribution of scalar gluon exchange \cite{Wang,Zhao:1998rt}.
	To date, the systematic researches of proton GFFs is very scarce and challenging. 
	The separate quark and gluon  EMT  operators
	are not conserved, for the additional form factors appear in the decompositions of their matrix elements. Moreover, the individual quark and gluon GFFs studies acquire scale- and scheme-dependence. 
	Therefore, we find it promising to obtain the complete information on proton GFFs from the light and heavy NTVMP process, which is essential for us to extract the related properties inside the proton.
	
	Even though the DVCS experiments is almost insensitive to gluon $D$-term, 
	the future experiments is crucial to gain insight into the proton's remaining GFFs, for instance, Jefferson Lab (JLab) \cite{Accardi:2023chb}, Electron-Ion Colliders (EICs) \cite{Accardi:2012qut,Burkert:2022hjz,Anderle:2021wcy}. 
	And this work will set important benchmarks for revealing the internal character inside the proton in future experiments at JLab and EICs. 
	
	\section{The internal nature of the proton}
	The expression of the GFFs is analogous to the electric charge distribution which can be mapped out by means of electron scattering experiments \cite{Fleurbaey:2018fih}. In a similar way, scattering off gravitons would allow one to access information on the spatial distribution of the energy inside a proton. 
	The physical content of the information encoded in the EMT form factors is revealed in the so-called Breit frame \cite{Polyakov:2018zvc}. 
	Here, the $(00)$ component of the EMT donate the isotropous energy density inside the proton, which can be expressed as \cite{Polyakov:2018zvc}
	\begin{align}\label{eq:T00r} 
		T_{00} (r)= \int \frac{d^{3} {\bf\Delta}}{(2 \pi)^{3}} e^{-i r {\bf\Delta} } G(t)
	\end{align}
	where $t=-{\bf\Delta}^2 $ 
	is the squared momentum transfer, the gravitational form factors $G(t)$ can be written as \cite{Polyakov:2018zvc,Ji:2021mtz,Wang:2022ndz}
	\begin{equation}\label{eq:GT} 
		G(t)= M A(t) - \frac{t}{4M} \left( A(t) -2J(t)+ D(t) \right)
	\end{equation}
	where $M$ is the proton mass. The form factors $A (t)$, $J (t) $ and $D (t)$ can derive the proton mass distribution, angular momentum distribution and mechanical properties, respectively. 
	Here $A(0)=1$, $J(0)=\frac{1}{2}$, and the form factor $B(t)=2J (t)-A (t)$ is consistent with zero basically \cite{Mamo:2021krl,Mamo:2019mka,Shanahan:2018pib}. 
	The proton GFFs are the sum contributions of the quark and gluon GFFs, which are given as \cite{Polyakov:2018zvc}
	\begin{equation} 
		G_{q/g}(t) = M A_{q/g}(t)- \frac{ t}{4 M }\left( -B_{q/g}(t)+D_{q/g}(t) \right) +M\bar{C}_{q/g}(t) 
	\end{equation} 
	here the $\bar{C}_{ g}(t) $ form factor can be written as \cite{Tong:2022zax,Wang:2022ndz}
	$$\bar{C}_{ g}(t)=-\frac{A_{ g}(t)}{4} + \frac{- t}{16 M^2 }\left( B_{ g}(t)-3 D_{ g}(t) \right) $$ 
	the $\bar{C} $-GFFs satisfies the constraint $\bar{C}_q(t) +\bar{C}_g(t) =0 $ due to EMT conservation \cite{Polyakov:2018zvc}. 
	As the contribution of the $B_{g}(t)+D_{g}(t)$ is negligible \cite{Tong:2022zax,Braun:2000kw,Braun:2006hz,Wang:2022ndz}, the component of the gluon GFFs part become \cite{Wang:2022ndz}
	\begin{equation}\label{eq:Gg} 
		\begin{aligned} 
			G_{g}(t) = & \frac{3}{4}M A_{g}(t) - \frac{ t}{4 M } D_{g}(t) +\frac{3t}{16M}\left(B_{g}(t)+D_{g}(t)\right) \\
			\approx & \frac{3}{4}M A_{g}(t) - \frac{ t}{4 M } D_{g}(t) 
		\end{aligned}
	\end{equation} 
	The quark GFFs are obtained as follows 
	\begin{equation}\label{eq:Gq} 
		\begin{aligned} 
			G_{q}(t) = & M A_{q}(t)- \frac{ t}{4 M }\left( -B_{q }(t)+D_{q }(t) \right) -M\bar{C}_{g}(t) \\
			\approx & M A_{q}(t)+ \frac{1}{4}M A_{g}(t) - \frac{ t}{4 M } D_{q}(t) -\frac{3t}{16M}\left( B_{g}(t)+D_{g}(t)\right) \\
			\approx & M A_{q}(t)+ \frac{1}{4}M A_{g}(t) - \frac{ t}{4 M } D_{q}(t) 
		\end{aligned}
	\end{equation} 
	as the form factor $B(t)=B_q(t)+B_g(t)\approx 0$.
	
	For $ A $-GFFs, the mass distribution of the proton are encoded in the $A$-term, which can be expressed under the dipole form parametrization as 
	\begin{align}
		A(t)=\frac{A_{q}(0)}{(1- t/m_q^{ 2} )^2}+\frac{A_{g }(0)}{(1- t/m_g^2 )^2}
	\end{align} 
	where the constraint $A_{q}(0)+A_{g }(0)=1$ is the consequence of momentum conservation \cite{Ji:1997pf,Polyakov:2018zvc}. $m_q$ and $m_g$ are free parameters. 
	The gluon contribution $A_g(0)=0.414$ was obtained from global QCD analysis \cite{Hou:2019efy}, 
	and agrees with other LQCD results \cite{Pefkou:2021fni,Alexandrou:2020sml,Yang:2018nqn}. In this work, to mirror the mass distribution inside the proton, a mean square radius of the mass distribution ($A$-term mass radius) is defined as
	\begin{align}\label{eq:RM}
		\left<R_{ A -term}^2\right>=\frac{\int d^{3} r r^{2} A(r)}{\int d^{3} r A(r)} = 12 \left(\frac{A_q(0)}{m_q^2} + \frac{ A_g(0)}{m_g^2} \right)
	\end{align}
	where $
	A (r)= \int \frac{d^{3} {\bf\Delta} }{(2 \pi)^{3}} e^{-i r {\bf\Delta} } A(t)
	$. 
	In previous work \cite{Kharzeev:2021qkd}, the expression $G(t)$ in Eq. (\ref{eq:GT}) is considerd as a mass distribution when $|t|$ is enough small as $G(t) \approx MA(t)$. Through the extraction of the form factor from the $J/\psi$ photoproduction at small $|t|$ \cite{Kharzeev:2021qkd,Wang:2022vhr,Duran:2022xag}, the result of considering gluon contribution can derive the gluon radius of the proton; while the radius derived from the $\rho^0$ differential cross section can be regarded as the quark contribution without the gluon part \cite{Wang}. 
	Actually, $A(t)=A_q(t)+A_g(t)$ denote the sum contribution inside the proton, which allow us to define the proton mass radius from $A$-term as Eq. (\ref{eq:RM}).
	
	The $D $-term in GFFs is an area of significant interest, and can define pressure and shear forces distributions inside the proton. The $D$-term is typically parameterized in the tripole form as \cite{Shanahan:2018nnv,Fiore:2021eav,Lepage:1980fj}
	\begin{align}\label{eq:Ggluon}
		D (t)=\frac{D_q(0)}{(1- t/ d_q^2 )^{3}}+\frac{D_g(0)}{(1-t/d_g^2)^{3}} 
	\end{align}
	where $D_q(0),d_q$ and $D_g(0),d_g$ correspond to the quark and gluon parameters, respectively.  
	
	As the gravity couples to matter via the EMT, the proton GFFs offers direct access to the proton's mechanical structure from the matrix elements of EMT \cite{Kobzarev:1962wt,Pagels:1966zza}. Concretely, the pressure $p(r)$ and shear forces $s(r)$ are “good observables” to report the pressure and shear forces distributions inside the proton, which can be expressed as \cite{Polyakov:2018zvc,Polyakov:2002yz}
	\begin{align}\label{eq:sr}
		s(r)= - \frac{1}{2} r\frac{d }{dr} \frac{1 }{ r} \frac{d }{dr} \widetilde{D} (r), ~~~~~~
		p(r)= \frac{1}{3} \frac{1 }{ r^2} \frac{d }{dr} r^2 \frac{d }{dr} \widetilde{D} (r)
	\end{align}
	here $\widetilde{D} (r)$ is the Fourier transform of $D(t)$ as \cite{Polyakov:2018zvc,Polyakov:2002yz}
	\begin{equation} 
		\widetilde{D} (r) =\int \frac{\text{d}^3 {\bf\Delta} }{2 M (2\pi)^3 } e^{-i {\bf\Delta} r}\left( D_{g}(- {\bf\Delta}^2) +D_{q}(- {\bf\Delta}^2) \right) 
	\end{equation} 
	
	The positive combination $ F_n(r)=\frac{2}{3} s(r)+p(r) $ has the meaning of the normal force distribution in the composed particle system \cite{Polyakov:2018zvc}. 
	One can define the proton mechanical radius in term of the meaning of the normal forces distribution in the system, which can be described as \cite{Polyakov:2018zvc}
	\begin{equation}\label{eq:mech} 
		\left<R^2_{mech}\right> = \frac{\int \text{d}^3r ~r^2 \left[F_n^{ q+ g }(r) \right] }{\int \text{d}^3r \left[F_n^{q+ g }(r) \right] } = \frac{ 12 \left( \frac{ D_q(0) }{ d_q}+ \frac{D_g(0)}{ d_g} \right) }{ D_q(0) d_q + D_g(0) d_g } 
	\end{equation}
	
	The energy density $T_{00} (r)$ in Eq. (\ref{eq:T00r}) can only be defined for the total system, which can be written as \cite{Polyakov:2018zvc}
	\begin{equation}\label{eq:T00g}
		\begin{aligned} 
			&T_{00}(r)= \int \frac{d^{3} {\bf\Delta} }{(2 \pi)^{3}} e^{-i r {\bf\Delta} } \left( G_q(t)+G_g(t) \right) \\
			&=\sum_{a=q,g}\frac{ 16 M^2 A_a(0) m_a^3 e^{-m_a r} - 
				(-3 + d_a r) D_a(0 ) d_a^5 e^{-d_a r} }{ 128 \pi M}
		\end{aligned}
	\end{equation}
	Moreover, the energy density satisfies $T_{00}(r)>0$ in a mechanical system, this allows us to introduce the mean square radius of the energy density (energy radius) as \cite{Polyakov:2018zvc}
	\begin{equation}\label{eq:RE} 
		\left\langle R_{E}^{2}\right\rangle =\frac{\int d^{3} r r^{2} T_{00}(r)}{\int d^{3} r T_{00}(r)} =\left<R_{ A -term}^2\right> -\frac{3\left(D_q(0)+D_g(0)\right)}{2M^2} 
	\end{equation}
	The proton energy radius can be associated with the slope of $A(t)$ in Eq. (\ref{eq:RM}) and the $D$-term GFFs, while covering the proton $A$-term mass radius. 
	
	\section{ The GFFs and vector meson photoproduction}
\begin{figure}[htbp]  
		\subfigure[ The gluon exchange mechanisms]{ 
			\includegraphics[width=0.44\textwidth]{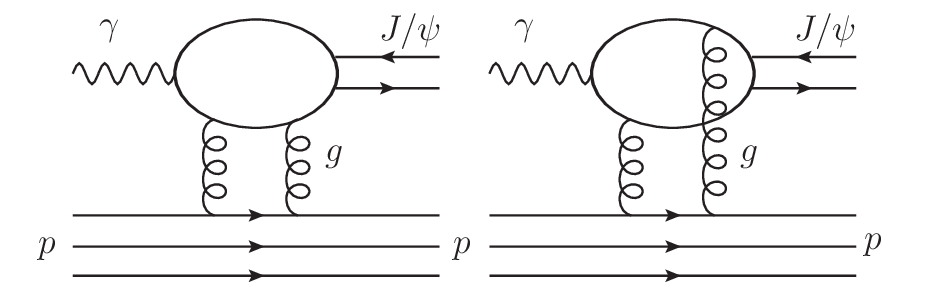} \label{gluon} 
		}
		\subfigure[ The  quark exchange mechanisms]{ 
			\includegraphics[width=0.44\textwidth]{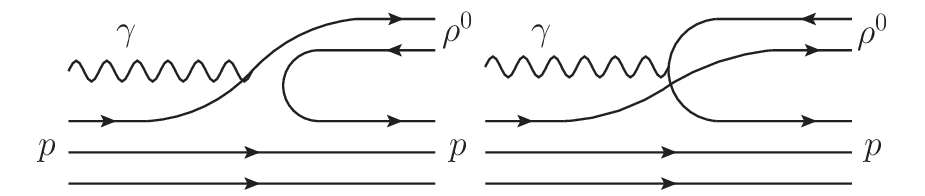} \label{quark} 
		}
		\caption{The Feynman diagrams for  $J/\psi$   production (  Fig.   \ref{gluon}) and  $\rho^0$ production  (Fig. \ref{quark}) in the $\gamma p\to V (\rho^0, J/\psi) p$  reaction. }
		\label{cross-section} 
	\end{figure}

	Next, we demonstrate the complete analysis of the gluon and quark GFFs from the near-threshold heavy quarkonium and light vector meson photoproduction, respectively. 
	For gluon GFFs, a recent work achieve great success  \cite{Duran:2022xag}, with the help of the pure nature of $J/\psi$ photoproduction, where the quark exchange is strongly suppressed by the Okubo-Zweig-Izuki (OZI) rule. Thus the heavy quarkonium photoproduction is a promising path to access the gluon GFFs due to the  gluon exchange dominant (Fig.   \ref{gluon}).   On the other hand,
	the quark GFFs can be natually associated with the $\rho^0$ photoproduction, as the   exchange between the proton and the meson states are occupied by valence quarks (Fig. \ref{quark}), %$\sigma, f_2, 2\pi$, and Pomeron
	and the gluon exchange is quite meager when $|t|$ is small \cite{CLAS:2001zxv,Wang,Zhao:1998rt}.

	Typically, the photoproduction cross section of the vector meson is given by \cite{ParticleDataGroup:2022pth}
	\begin{align}\label{eq:differential}
		\frac{d\sigma_{\gamma p\to V p}}{dt}=\frac{1}{64 \pi W^2} \frac{1}{|{\bf p}_{\gamma}|^2} \left| \mathcal{M}_{\gamma p\to V p} \right|^2
	\end{align}
	where ${\bf p}_{\gamma }$ and $W$ are the center of mass (c.m.) photon momentum and c.m. energy in the $ \gamma p\to V p$ process, respectively. As an assumption, the amplitude of light and heavy quarkonium primarily attributes to the quark and gluon part
	of the EMT of QCD in this work, respectively, which can be written as 
	\cite{Kharzeev:2021qkd} 
	\begin{align}\label{eq:amplitude1} 
		\mathcal{M}_{\gamma p\to V p} =-Q _e c_2 2 M g^2 \left< P^{\prime}| T^{q(g)}_{00} |P \right> 
	\end{align}
	where $Q_e $ represents the coupling of the photon to the electric charge of the quarks in vector meson. 
	$ c_2 $ is the short-distance coefficient, and
	$g^2=4$ is the QCD coupling with $\alpha_s \approx 0.32 $ \cite{Hatta:2018ina}. 
	Combining Eqs. (\ref{eq:Gg}) and (\ref{eq:Gq}), the component of the quark and gluon part EMT form factors become 
	\begin{equation}\label{eq:G00} 
		\begin{aligned} 
			&\left< P^{\prime}|T^{g }_{00}|P \right> =\bar{u} (P^{\prime}) u(P) \left(\frac{3}{4} M A_{g }(t) - \frac{ t}{4 M } D_{ g }(t) \right) \\
			& \left< P^{\prime}|T^{q}_{00}|P \right> =\bar{u} (P^{\prime}) u(P) \left( M A_{q}(t)+ \frac{1}{4}M A_{g}(t) - \frac{ t}{4 M } D_{q}(t) \right)
		\end{aligned} 
	\end{equation} 
	where the spinor normalization $\bar{u} (P^{\prime}) u(P)=2M$. 
	By integrating the differential cross section (Eq. (\ref{eq:differential}))
	over the allowed kinematical range from $t_{min}$ to $ t_{max}$, 
	the total cross section are computed and can be expressed as 
	\begin{align}\label{eq:total} 
		\sigma=\int_{t_{min}}^{t_{max}} dt \left(\frac{d \sigma}{d t} \right) ,
	\end{align}

	It has been determined that the form factor $G_{q/g}(t)$ in Eqs. (\ref{eq:Gg}) and (\ref{eq:Gq}) satisfy
	\begin{align}
		G_q(0)= MA_q(0)+\frac{1}{4}MA_g(0)\text{~~and~~} 
		G_g(0)=\frac{3}{4}MA_g(0)
	\end{align} 
	whereby, the coefficient $ c_2 $ in Eq. (\ref{eq:amplitude1}) can be determined by extrapolating the near-threshold differential cross section at $t=0$. 
	So far, we have established the relationship between the complete GFFs and the NTVMP,  including the differential and total cross section in Eqs. (\ref{eq:differential}) and (\ref{eq:total}). 
	
\begin{figure}[htbp] 
		\includegraphics[width=0.45\textwidth]{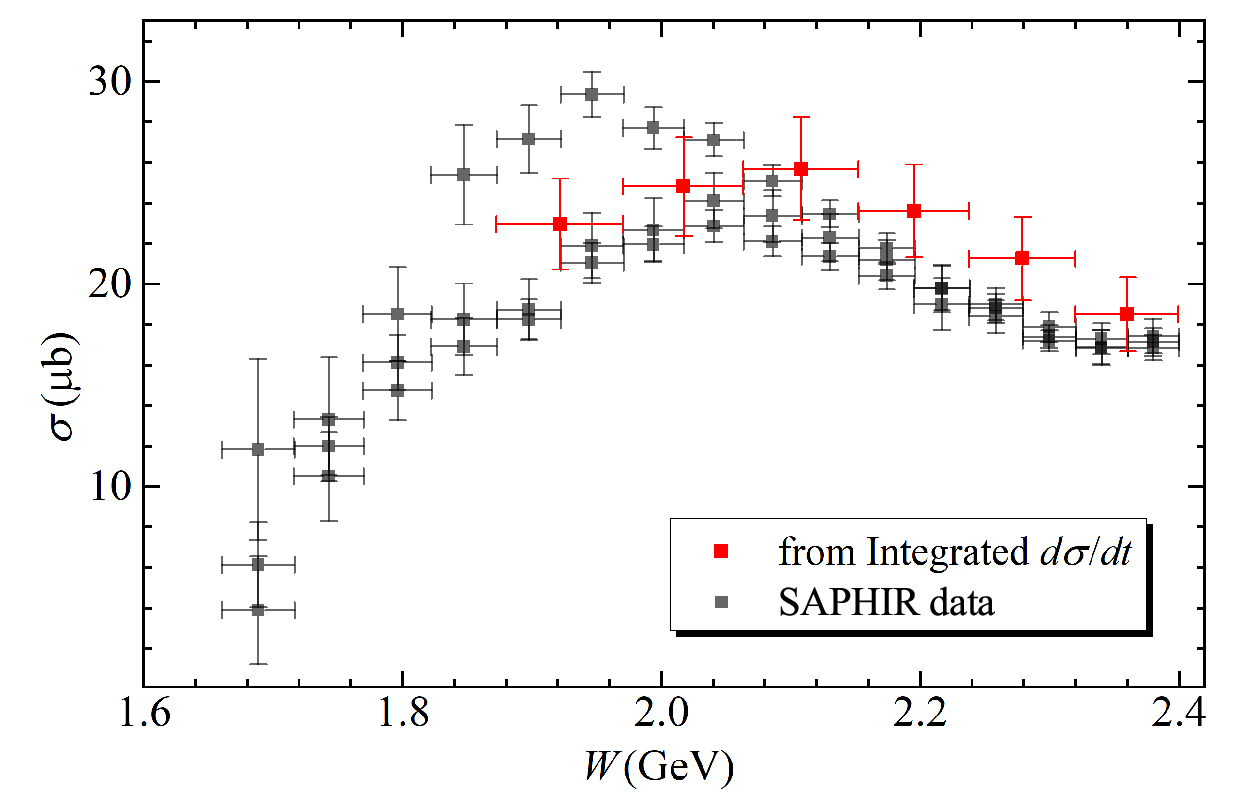}
		\caption{ The total cross sections calculated by integrating the functions
			fitted to the differential cross sections for the six
			beam energy regions, with the statistical uncertainties of $m_q$, $D_q(0)$ and $ d_q$ shown, and compared with the experimental results (black squares) \cite{Wu:2005wf}.  } 
		\label{total-cross-section} 
	\end{figure}

	As discussed in Ref. \cite{Wang:2022ndz}, the parameters $m_g$, $D_g(0)$ and $d_g$ in gluon GFFs are attained by global fitting the near-threshold $J/\psi$ photoproduction data,  including the differential and total cross section  
	\cite{Duran:2022xag,GlueX:2019mkq,Gittelman:1975ix,GlueX:2019mkq,Camerini:1975cy,Frabetti:1993ux,Amarian:1999pi}.
	The quark GFFs calculation are resemblance to the gluon GFFs (as described in Ref. \cite{Wang:2022ndz}). We consider the squared momentum transfer  $t$ of the differential cross section to be in the range of   $|t|<1$ GeV$^2$. 
	A meticulously global fitting of the $\rho^0$ photoproduction experimental data is employed \cite{Wu:2005wf}, to extract the parameters $m_q$, $D_q(0)$ and $ d_q$ in quark GFFs ( Eqs. (\ref{eq:differential})). These obtained parameters as presented in Table \ref{tab:D}. The comparison between the differential cross section of $\rho^0$ photoproduction and the experimental measurements is manifested in Fig. \ref{cross-section}, exhibiting a good agreement. 
	To check the accuracy of the differential cross sections fitting, we compare these integrals vulue through Eq. (\ref{eq:total}) with the total cross section experimental data \cite{Wu:2005wf}, as depicted in Fig. \ref{total-cross-section}. 
	Our result demonstrates agreement with the value and trend of experimental data, which has
	proven the reliable fitting.

	\begin{figure}[htbp] 
		\includegraphics[width=0.47\textwidth]{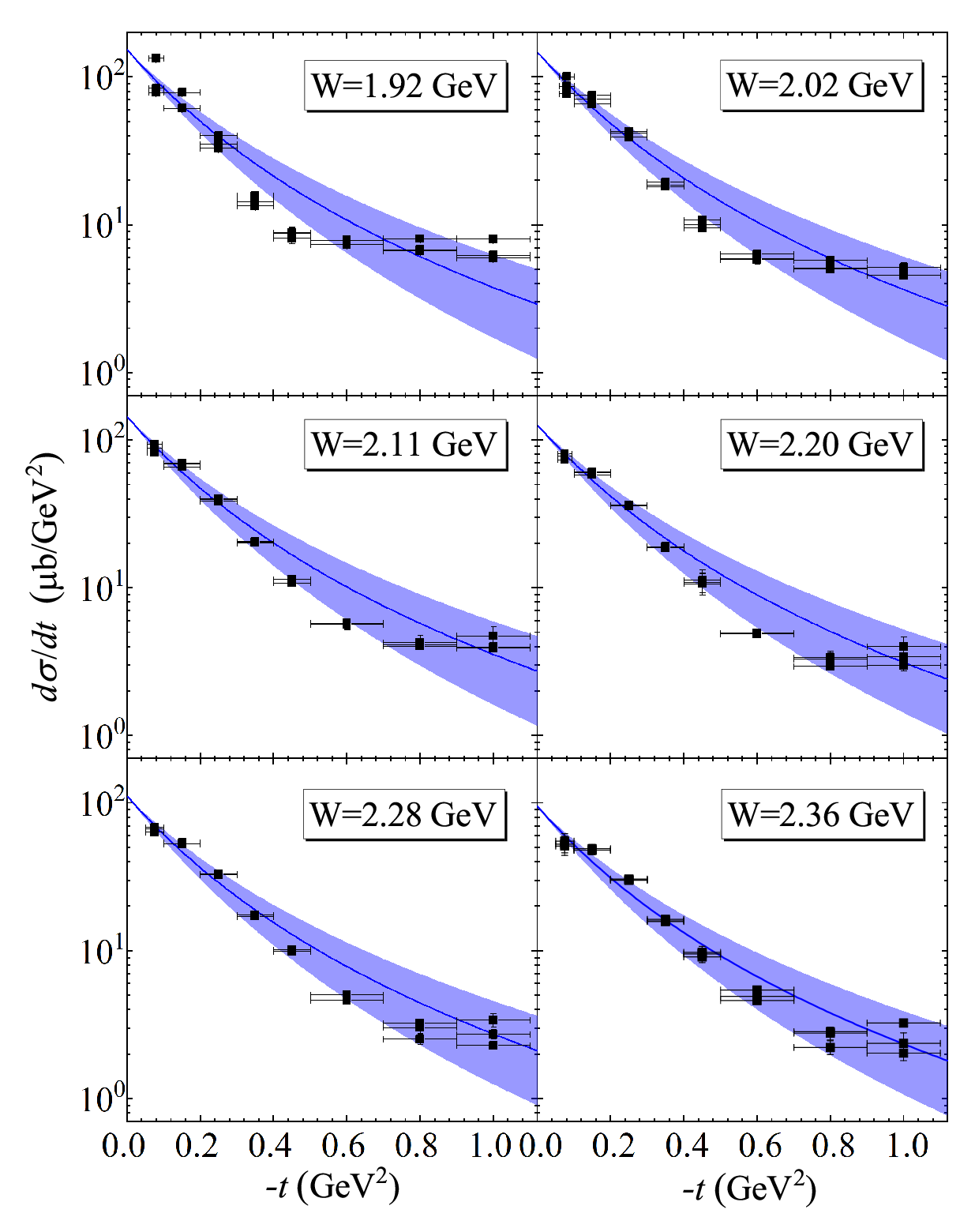}
		\caption{ Global fitting result of $\gamma p \to \rho^0 p$ differential cross section as a function of $-t$ at c.m. energy $1.92 \text{ GeV},$ $ 2.02 \text{ GeV}, $ $2.11 \text{ GeV}, $ $2.20 \text{ GeV},$ $ 2.28 \text{ GeV} $, and $2.36\text{ GeV}$. The blue bands reflect  the statistical error of $m_q$, $D_q(0)$ and $ d_q$. Reference of data can be found in \cite{Wu:2005wf}. } 
		\label{cross-section} 
	\end{figure}

	\begin{table}
		\caption{Top: obtained values of the parameters $m_q$, $D_q(0)$ and $ d_q$ in quark GFFs by a global fitting of the $\rho^0$ differential cross section experimental data \cite{Wu:2005wf}. Bottom: the parameters $m_g$, $D_g(0) $ and $d_g$ in gluon GFFs achieved from global fitting the $J/\psi$ photoproduction data \cite{Duran:2022xag,GlueX:2019mkq,Gittelman:1975ix,GlueX:2019mkq,Camerini:1975cy,Frabetti:1993ux,Amarian:1999pi}. } 
		\begin{tabular}{c cccccc }
			\hline\hline\noalign{\smallskip}
			\multirow{2}*{~quark GFFs (this work)~} & $m_q$ (GeV) & $D_q(0)$ & $ d_q$ (GeV) \\ \noalign{\smallskip} 
			& $0.85\pm 0.05$ & $-1.76\pm 0.30$ & $0.88\pm 0.05$ \\ 
			\hline \noalign{\smallskip}
			\multirow{2}*{~~gluon GFFs \cite{Wang:2022ndz}~~} & $m_g$ (GeV) & $D_g(0)$ & $ d_g$ (GeV) \\ \noalign{\smallskip} 
			& ~$1.43\pm 0.10$~ & ~$-2.08 \pm 0.25$~ & ~$0.90 \pm 0.07$ ~\\ 
			\hline\hline
		\end{tabular} 
		\label{tab:D}
	\end{table}
	
\begin{figure}[htbp]
		\centering 
		\includegraphics[width=0.4\textwidth]{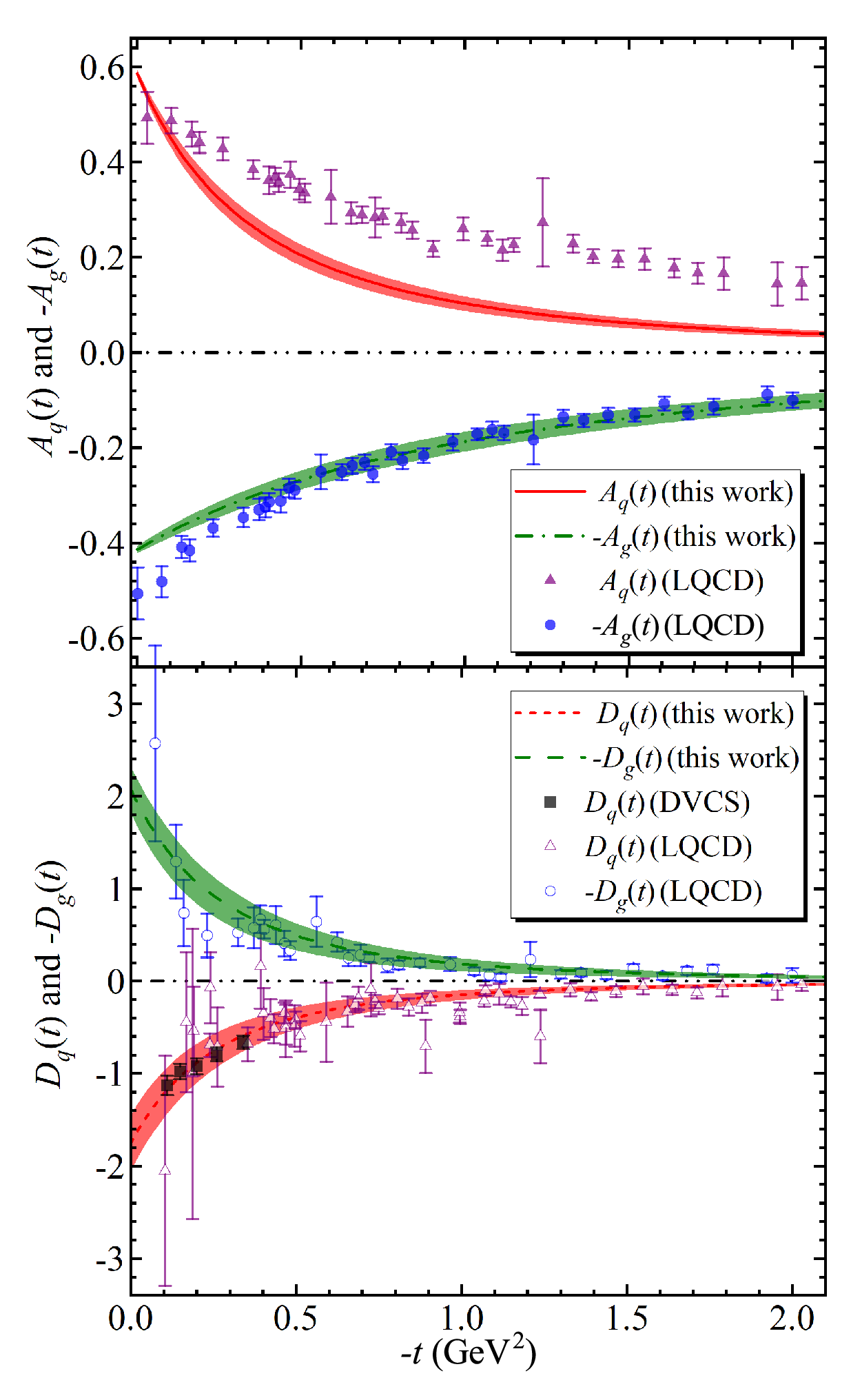}
		\caption{Top panel: The $A_{q}(t)$ (red solid curve) and converse $A_{g}(t)$ (green dash-dotted curve) compared with LQCD results \cite{Hackett:2023rif}.
			Bottom panel: The $D_{q}(t)$ (red dotted curve) and converse $D_{g}(t)$ (green dashed curve) 
			compared with LQCD results \cite{Hackett:2023rif} and the result extracted from DVCS experiments \cite{Burkert:2018bqq}. } 
		\label{DqDg} 
	\end{figure}

\begin{figure}[htbp]
		\centering 
		\includegraphics[width=0.4\textwidth]{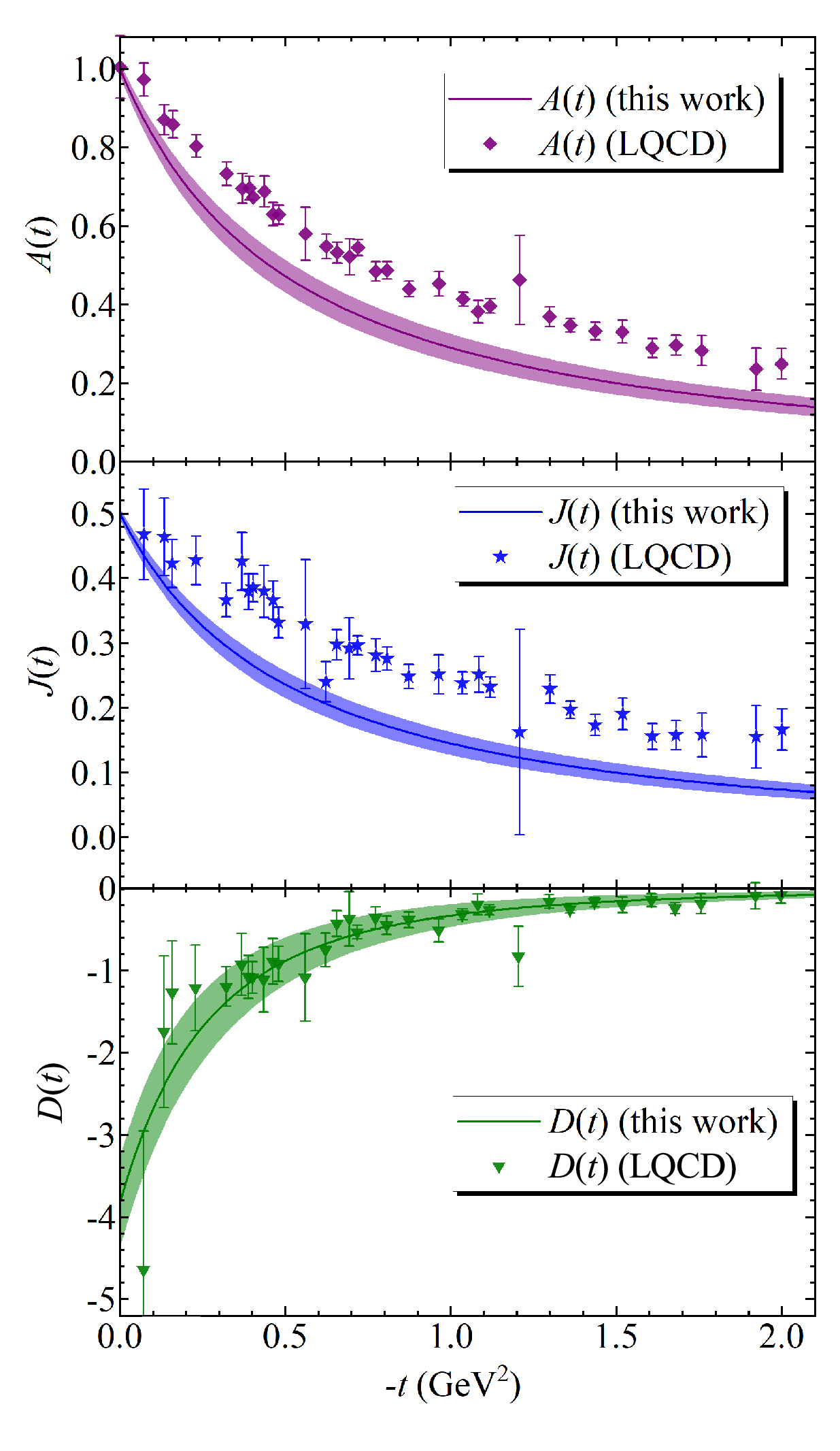}
		\caption{The three GFFs of the proton, including $A(t)$,$J(t)$,and $D(t)$, as a function of $t$, are compared with LQCD results \cite{Hackett:2023rif}.
		} 
		\label{total-GFFs} 
	\end{figure}

	As shown in Fig. \ref{DqDg}, the value of quark and gluon $A$-term and $D$-term are compared with the DVCS experiment data and LQCD determinations \cite{Hackett:2023rif}. Here the errors of parameters include all uncertainties of the GFFs. 
	For quark GFFs, the quark $D$-term exhibits a notable agreement with the DVCS experiment data and LQCD determinations, while our statement on $A$-term is approximately half smaller than the LQCD results \cite{Hackett:2023rif}.
	The converse gluon $A$- and $D$-term are displayed in Fig. \ref{DqDg}, to avoid the overlapping of the data points. Notably, the gluon $D$-term contributes to more than the quark $D$-term lightly, and our statement obtained for gluon $A$ and $D$-term are approximately comparable with the LQCD results \cite{Hackett:2023rif}.
	The total GFFs are shown in Fig. \ref{total-GFFs}, compared with the LQCD  results \cite{Hackett:2023rif}. For the $D(t)$, our results are found to be consistent with LQCD result, while both $A(t),J(t)$ is a slightly small.
	The deviation of $A(t)$ can be attributed to the absence of $s$ quark contribution in our calculations, then the $J(t)=\frac{1}{2}A(t)$ can be derived. Therefore,  further follow-up sufficient work is still necessary.
	
	The proton $A$-term mass radius, mechanical radius and energy radius in Eq. (\ref{eq:RM}), (\ref{eq:mech}) and (\ref{eq:RE}) are computed, as listed in Table \ref{tab:radius}.   
	Two important mechanical quantity,   the pressure  and   energy density  in the center proton, are also revealed. 
	Astonishingly, the obtained proton $A$-term mass radius is a notable achievement that the corresponding ratio of the radius derived from mass and electromagnetic form factor is $0.82\pm0.05$, which approach to the $\pi$ meson's determination 0.79(3) \cite{Xu:2023bwv}.
	It turns out that the ratio of the mass radius and charge radius of different hadrons seems to be consistent. 
	
	\begin{figure}[htbp]
		\centering 
		\includegraphics[width=0.4\textwidth]{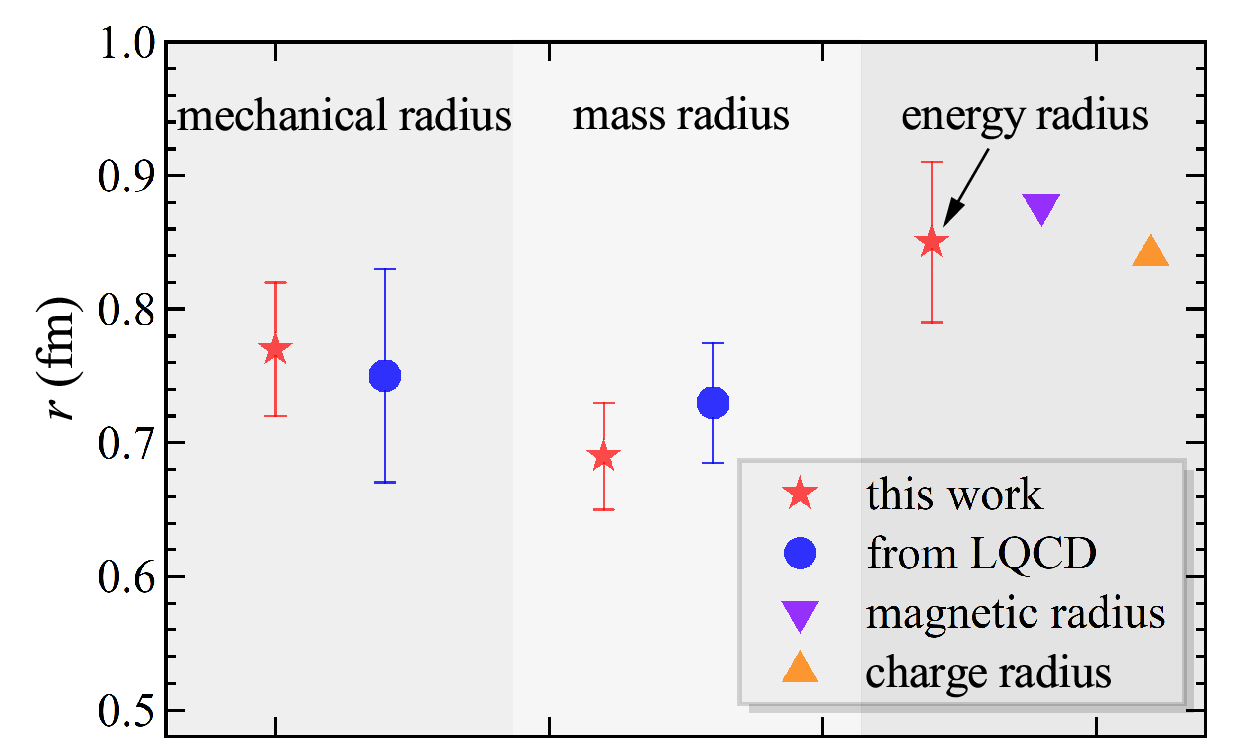}
		\caption{Cimparison of different proton radius, including the $A$-term  mass radius, mechanical radius and energy radius from this work and LQCD \cite{Hackett:2023rif}. The proton charge and magnetic radius are also shown \cite{Bernauer:2020ont}. } 
		\label{radius} 
	\end{figure}

	Notably, the energy radius derived from GFFs in Eq. (\ref{eq:RE}) can also be written as 
	\begin{equation}\label{eq:energy2}
		\left\langle R_{E}^{2}\right\rangle=\frac{\int d^{3} r r^{2} T_{00}(r)}{\int d^{3} r T_{00}(r)}=\left.-\frac{6}{G(0)} \frac{d G(t)}{dt}\right|_{t=0} 
	\end{equation}
	which bears a striking resemblance to proton charge and magnetic radius defined from the electric and magnetic form factor \cite{Bernauer:2020ont}.
	Astonishingly, the proton energy radius $ 0.85\pm 0.06 \text{ fm}$ is close in size to its charge and magnetic radius, as shown in Fig. \ref{radius}. The proton $A$-term mass radius and mechanical radius are also compared.
	It reveals an intriguing pattern that the electric, magnetic, as well as gravitational form factor is homologous and delicately governs the
	enigmatic properties of the proton. Therefore, the accurate measurement
	of the proton energy radius  deserves the same attention as charge and magnetic radius in the future study.

	\begin{table} 
		\caption{The extracted internal nature of the proton from the GFFs: $A$-term mass radius $\sqrt{\left<R_{ A -term}^2\right> }$, mechanical radius $\sqrt{\left<R^2_{mech}\right> }$, energy radius $\sqrt{\left\langle R_{E}^{2}\right\rangle }$, the pressure and energy density in the center of the proton. } 
		\begin{tabular}{c cccccc }
			\hline\hline\noalign{\smallskip}
			$A$-term mass radius (fm) & mechanical radius (fm) & energy radius (fm) \\ \noalign{\smallskip}\hline\noalign{\smallskip}
			$0.69 \pm 0.04$ & $0.77\pm 0.05$ & $ 0.85\pm 0.06$ \\ 
			\noalign{\smallskip}\hline\noalign{\smallskip} 
			$p(0)$ (GeV/fm$^3$) & $T_{00}(0)$ (GeV/fm$^3$) &--- \\ \noalign{\smallskip}\cline{1-3}\noalign{\smallskip}
			$ 1.49^{+0.87}_{-0.60}$ & $5.40^{+2.54}_{-2.76}$  &--- \\ 
			\noalign{\smallskip}
			\cline{1-3}
			\cline{1-3}
		\end{tabular} 
		\label{tab:radius}
	\end{table}
	
\begin{figure}[htbp]
		\centering 
		\includegraphics[width=0.5\textwidth]{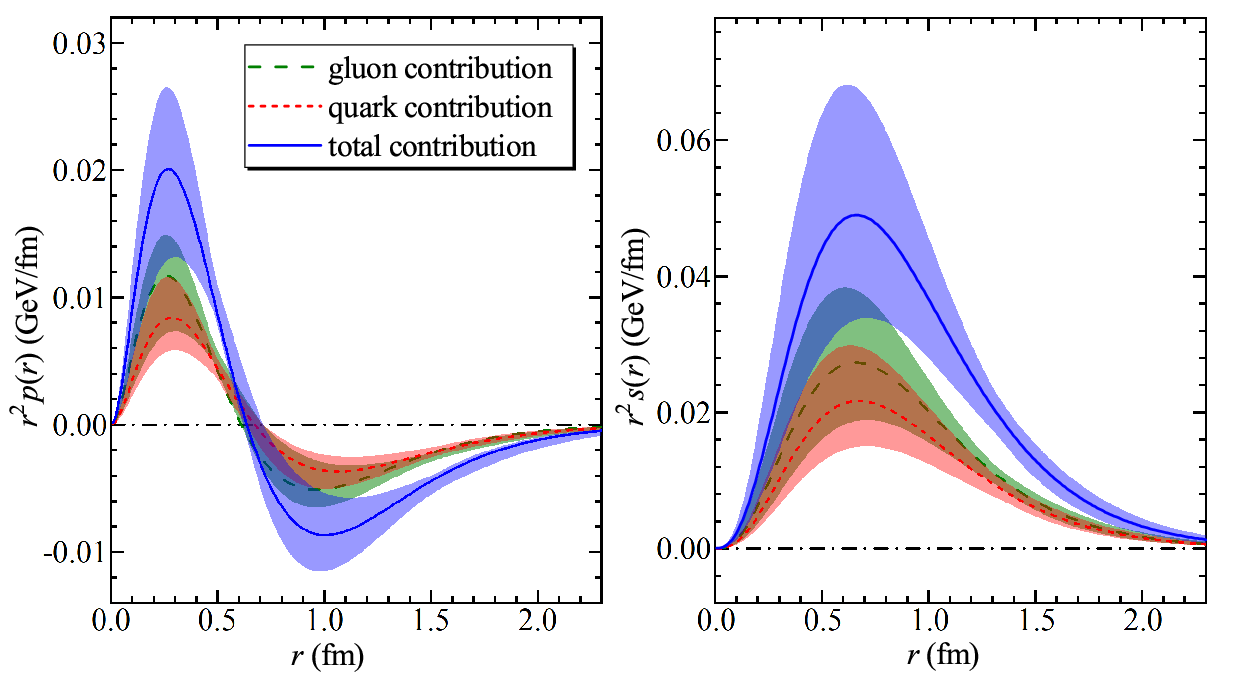}
		\caption{Left panel: the pressure distribution $r^2p(r)$ of the proton. Right panel: the shear forces distribution $r^2s(r)$ of the proton. The blue solid curve corresponds to the total pressure and shear forces distribution, while the red dotted and green dashed curve denote the quark and gluon contributions to the total, respectively. } 
		\label{sp-pr} 
	\end{figure}

	The pressure and shear forces distributions inside the proton are obtained and displayed in Fig. \ref{sp-pr}. 
	Here the positive sign means repulsion towards the outside, and the negative sign indicates attraction directed towards the inside. 
	The total pressure and shear forces distributions of the sum of the quark and gluon contributions are illustrated as the blue-solid curve in Fig. \ref{sp-pr}.
	It was found that the 
	pressure is positive in the inner region, and negative in the outer region, 
	with a zero crossing near $ r $=$ 0.63 \text{ fm}$, which shows the repulsive and binding pressures dominate in the proton are separated in radial space. 
	Moreover, the shear forces reaches its distribution peaking near $r$=0.66 fm in our observation.

	\section{$D_q(t)$ result with model-independent algorithm}
	In our previous work, $D_q(t)$ results for the NTVMP process were given. However, comparison with the results of other models in Ref.\cite{Polyakov:2018zvc}  that there are large differences between the different models. Therefore we introduce a generalized model-independent algorithm for further validation, which is called the Resonances Via Pad\'{e} (RVP) method based on the Schlessinger Point Method (SPM) \cite{Schlessinger:1966zz,Schlessinger:1968vsk,Tripolt:2016cya}. The method is essentially an interpolation approximation that relies only on experimental data points. When given a finite dataset $\{(x_i,y_i)|_{i=1,2,\cdots,N}\}$ with a data volume of $N$, the function $y(x)$ can be accurately reconstructed by constructing the interpolating continuous partition $C_N(x)$ via SPM.
	\begin{equation}
		C_{N}(x)=\frac{y_{1}}{1+\frac{a_{1}\left(x-x_{1}\right)}{1+\frac{a_{2}\left(x-x_{2}\right)}{\vdots_{a_{N-1}\left(x-x_{N-1}\right)}}}}
		\label{s1}
	\end{equation}
	where the coefficients $\{a_i|i=1,2,\cdots,N-1\}$ can be constructed recursively via Eq. (\ref{s2}) and ensured that $y(x_i)=C_N(x_i)$.
	\begin{equation}
		\begin{aligned}
			&a_{1}=\frac{y_{1} / y_{2}-1}{x_{2}-x_{1}} \\ &a_{i}=\frac{1}{x_{i}-x_{i+1}}\left[1+\frac{a_{i-1}\left(x_{i+1}-x_{i-1}\right)}{1+} \cdots \frac{a_{1}\left(x_{i+1}-x_{1}\right)}{1-y_{1} / y_{i+1}}\right] 
		\end{aligned}
		\label{s2}
	\end{equation}

	For GFFs, only experimental data from DVCS were available. Therefore the form of $D_q(t)$ under the RVP method can be obtained by bringing the experimental data points from DVCS into Eqs. (\ref{s1})  (\ref{s2}) and simplifying it:
	\begin{equation}
		D_q^{\text{RVP}}(Q)=\frac{0.061913 - 0.346966 Q - 0.004908 Q^2}{-0.033875 + 0.010085 Q +  Q^2}
		\label{s3}
	\end{equation}
	where $Q=-t$.
	\begin{figure}[htbp]
		\centering 
		\includegraphics[width=0.45\textwidth]{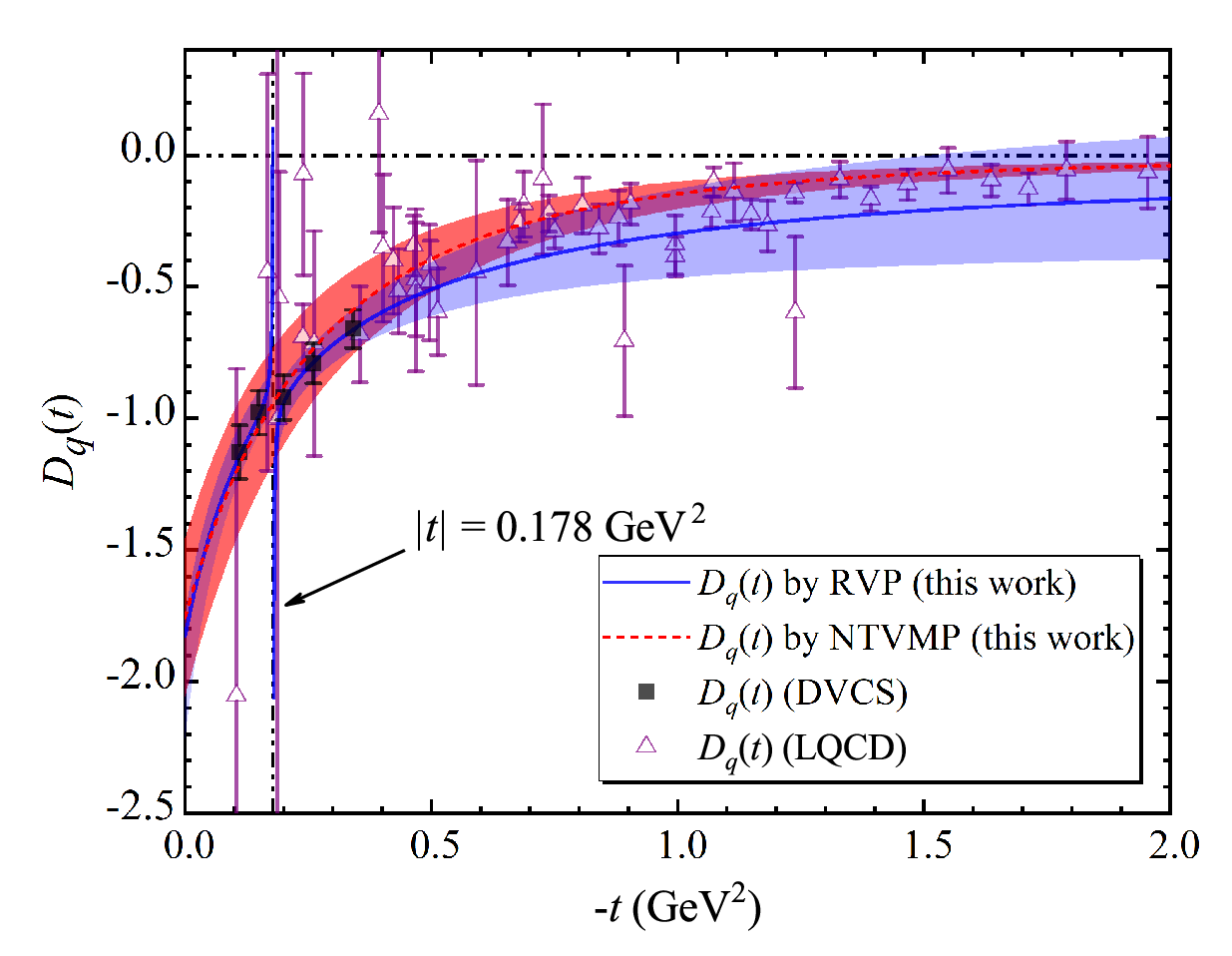}
		\caption{Blue solid line: quark $D$-term  results obtained by reconstructing the DVCS experimental data points using the RVP method based on SPM; Red dashed line: quark $D$-term in NTMVP process; And compared with the LQCD results.} 
		\label{SPM} 
	\end{figure}

	The comparison of the numerical results of Eq. (\ref{s3}) with the NTVMP process is shown in Fig. \ref{SPM}. It can be seen that the result of $D_q(t)$ of this model-independent algorithm is more supportive of NTVMP process. In addition, comparing to other models in Ref. \cite{Polyakov:2018zvc}, the results of these works seem to conform better to the experimental data and LQCD determinations. However, there are several issues that need to be clarified regarding the results obtained by the RVP method:

	\begin{itemize}
		\item Feasibility of interpolation and extrapolation : The RVP method based on the SPM belongs to a kind of Pad\'{e} approximation, which can accurately reconstruct an underlying function from data points. Its radius of convergence is determined by a branch point of the function, which is located near the real-axis domain of the data sample. Also the discrete nature of the method ensures that its results can be reasonably extended over a wider range. For example, $|t|\in [0.1 , 0.4] \text{GeV}^2$ in the experimental data of DVCS, whereas we finally extrapolate our results to $|t| = 2  \text{GeV}^2$ and are in good agreement with both NTVMP process and LQCD determinations. In addition, other hadron physics related problems are accurately calculated and their feasibility is illustrated in Ref. \cite{Tripolt:2016cya,Binosi:2022ydc,Cui:2021vgm,Cui:2021skn} by this method.

		\item Identification of threshold energies and poles:  From the results in Fig. \ref{SPM}, it can be noticed that the results shown by the RVP method have a pole at $|t| = 0.178 \text{GeV}^2$ and that the point also corresponds to the result with the largest error in the LQCD. This may be a coincidence, but it was pointed out in Ref. \cite{Tripolt:2016cya,Binosi:2022ydc} that the RVP method has the ability to identify threshold energies and poles. Because the underlying function reconstructed by this method exhibits non-canonicality as it approaches the poles, and this non-canonical behaviour provides a clear indication of the region where the poles are located. Therefore, we conjecture that there may be some kind of interesting physics phenomenology at  $|t| = 0.178 \text{GeV}^2$, which we will investigate in our subsequent work.
	\end{itemize}
	
	At present, due to the limited experimental data, we only calculate the quark $D$-term in GFFs by RVP method. In the future, with the updating of experimental data, we will use this method to systematically study the GFFs of proton more deeply.
	\section{ SUMMARY AND DISCUSSION} 
	In this work, based on the present light vector meson $\rho^0$ and heavy quarkonium $J/\psi$ photoproduction data, a groundbreaking analysis of the complete proton GFFs connecting the energy-momentum tensor and the NTVMP are investigated. 
	The obtained GFFs are compared with the DVCS experimental results and LQCD determinations \cite{Hackett:2023rif}. 
	It turns out that the quark $D$-term is comparable with the current result, while differences persist between our statements on the quark $A$-term and the LQCD results \cite{Hackett:2023rif}. 
	For gluon GFFs, the value of gluon $A$- and $D$-term are approximately consistent with the LQCD results \cite{Hackett:2023rif}. 
	These results presented within this work illustrate the remarkable revelation that the NTVMP process offer tremendous potential for obtaining the complete proton GFFs. 
	Conversely, the connection between heavy quarkonium photoproduction and gluon GFFs also faces challenges. One study revealed that this process is light-cone dominated and has no direct connection with gluon GFFs \cite{Sun:2021pyw}. Thereby, further exploration of the physical mechanisms becomes imperative.

	This work employ a first-ever definition pattern about the mean square radius of the mass distribution, %that the $A$-term mass radius is definited from $A$-term
	comprising the sum contributions of quark and gluon inside the proton. 
	This  definition pattern in Eq. \ref{eq:RM} differs from that of the previous works.
	From   our perspective, the previous direct extraction from vector meson photoproduction \cite{Kharzeev:2021qkd,Wang:2022vhr,Duran:2022xag,Wang}  reflects the energy distribution (see Eqs. (\ref{eq:RE}) and (\ref{eq:energy2})) instead of the mass distribution inside the proton. Because these works assume that $G(t)$ can be considered  as a mass distribution when $|t|$ is small enough that $G(t) \approx MA(t)$. This study tries to clarify this point and develops the specific values of the $A$-term mass radius and energy radius from the   mass and energy distribution.
	Moreover, the total $ D$-form factor is helpful for investigating their applications in descripting the mechanical properties. Correspondingly, the pressure and shear forces distributions inside proton are obtained.  We achieve the internal nature of the proton, including the proton $A$-term mass radius, mechanical radius and energy radius, normal forces and energy density distribution in the proton system, respectively. 
	
	Considering that there are differences in the distributions of GFFs of proton given by different theoretical models \cite{Polyakov:2018zvc}. In order to reduce the influence of parameters and models on physical results, we use RVP method to compute quark $D$-term based on DVCS experimental data . The results demonstrate that this model-independent generalized algorithm is more supportive of the NTVMP process in this work.  Subsequently, we also consider other computational methods, such as machine learning \cite{Tripolt:2016cya,Binosi:2022ydc,Cui:2021vgm,Cui:2021skn,Dutrieux:2021nlz}, to do further research targeting other contributions of GFFs. 
	%Additionally,  these  new  approaches  can  investigate  the model  outcomes  for $d\sigma /dt|_{t=0}$.    
	
	In this work, the NTVMP process offer significant information for the study of proton GFFs.
	Therefore, the high-precision photo/electroproduction data of light and heavy vector mesons (including $\rho^0, \omega$, $ \phi, J/\psi,$ and $ \Upsilon$) are very  needed, which can be realized in the JLab \cite{Accardi:2023chb} and EICs \cite{Accardi:2012qut,Burkert:2022hjz,Anderle:2021wcy}. 
	Moreover, this study will offer theoretical reference for the   GFFs measurement in the future DVCS experiments and other  indirect measurement in exclusive processes.

	\section*{Acknowledgments}
	
	This work is supported by the National Natural Science Foundation of China under Grant Nos. 12065014 and 12047501, and by the Natural Science Foundation of Gansu province under Grant No. 22JR5RA266. We acknowledge the West Light Foundation of The Chinese Academy of Sciences, Grant No. 21JR7RA201.

\end{document}